\documentclass[journal=ancac3,manuscript=article]{achemso} %
\usepackage{graphicx}
\usepackage{fancyhdr}
\usepackage{xcolor}
\usepackage{makeidx}
\usepackage{amssymb,amsmath}
\usepackage{physics}
\usepackage{upquote}
\usepackage{array}
\usepackage{enumerate}
\usepackage{rotating}
\usepackage{tabularx}
\usepackage{adjustbox}
\usepackage{array}
\usepackage{cite}
\usepackage{multirow}
\usepackage{abstract}
\usepackage{natbib}
\usepackage{pdfpages}

\newcommand{\bc}{\begin{center}}
\newcommand{\ec}{\end{center}}

\newcommand{\beq}{\begin{equation}}
\newcommand{\eeq}{\end{equation}}
\newcommand{\beqs}{\begin{eqn*}}
\newcommand{\eeqs}{\end{eqn*}}
\newcommand{\bq}{\begin{quote}}
\newcommand{\eq}{\end{quote}}

\newcommand{\ben}{\begin{enumerate}}
\newcommand{\een}{\end{enumerate}}
\newcommand{\bit}{\begin{itemize}}
\newcommand{\eit}{\end{itemize}}

\newcommand{\ba}{\begin{array}}
\newcommand{\ea}{\end{array}}
\newcommand{\beqa}{\begin{eqnarray}}
\newcommand{\eeqa}{\end{eqnarray}}
\newcommand{\beqas}{\begin{eqnarray*}}
\newcommand{\eeqas}{\end{eqnarray*}}
\newcommand{\bfg}{\begin{figure}}
\newcommand{\efg}{\end{figure}}

\newcounter{algc}
\newcounter{romc}
\newcounter{Alphc}
\newcommand{\bl}{\begin{list}{{\it Step} ~\arabic{algc}~:} {\usecounter{algc}
			\setlength{\topsep}{0pt} \setlength{\itemsep}{0pt}}}
	\newcommand{\el}{\end{list}}
\newcommand{\blr}{\begin{list}{~\roman{romc}~:} {\usecounter{romc}
			\setlength{\topsep}{0pt} \setlength{\itemsep}{0pt}}}
	\newcommand{\elr}{\end{list}}
\newcommand{\bla}{\begin{list}{~\Alph{Alphc}~:} {\usecounter{Alphc}
			\setlength{\topsep}{0pt} \setlength{\itemsep}{0pt}}}
	\newcommand{\ela}{\end{list}}
\newcommand{\tsup}{\textsuperscript}
\newcommand{\tsub}{\textsubscript}
\newcolumntype{L}{>{\centering\arraybackslash}m{3cm}}
\newenvironment{conditions}
{\par\vspace{\abovedisplayskip}\noindent\begin{tabular}{>{$}l<{$} @{${}={}$} l}}
	{\end{tabular}\par\vspace{\belowdisplayskip}}
\author{Mehak Mahajan}
\author{Kausik Majumdar}
\email{kausikm@iisc.ac.in}
\affiliation[Indian Institute of Science]
{Department of Electrical Communication Engineering, Indian Institute of Science, Bangalore 560012, India}
\title{Gate- and Light-Tunable Negative Differential Resistance with High Peak Current Density in 1T-TaS\tsub2/2H-MoS\tsub2 T-Junction}
\abbreviations{NDR}
\begin{document}
\begin{tocentry}
\centering	
\includegraphics[]{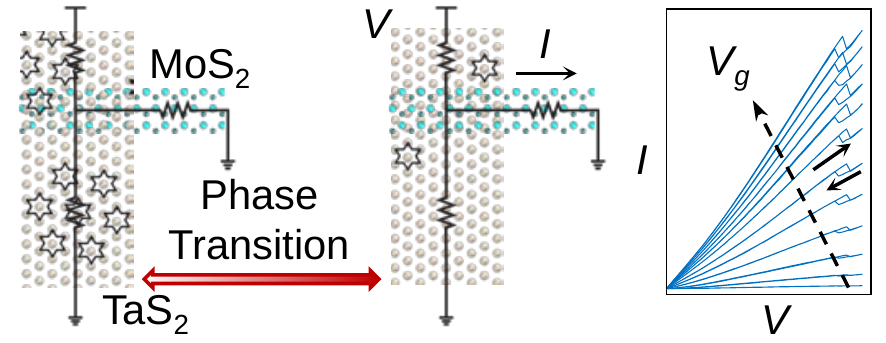}
\end{tocentry}
\begin{abstract}
Metal-based electronics is attractive for fast and radiation-hard electronic circuits and remains one of the longstanding goals for researchers. The emergence of 1T-TaS\tsub2, a layered material exhibiting strong charge density wave (CDW) driven resistivity switching that can be controlled by an external stimulus such as electric field and optical pulses, has triggered a renewed interest in metal-electronics. Here we demonstrate a negative differential resistor (NDR) using electrically driven CDW phase transition in an asymmetrically designed T-junction made up of 1T-TaS\tsub2/2H-MoS\tsub2 van der Waals heterostructure. The principle of operation of the proposed device is governed by majority carrier transport and is distinct from usual NDR devices employing tunneling of carriers, thus avoids the bottleneck of weak tunneling efficiency in van der Waals heterojunctions. Consequently, we achieve a peak current density in excess of $10^5$ nA$\mu$m$^{-2}$, which is about two orders of magnitude higher than that obtained in typical layered material based NDR implementations. The peak current density can be effectively tuned by an external gate voltage as well as photo-gating. The device is robust against ambiance-induced degradation and the characteristics repeat in multiple measurements over a period of more than a month. The findings are attractive for the implementation of active metal-based functional circuits.
\end{abstract}

\newpage
Metal-based electronics remains one of the longstanding goals of researchers to achieve ultra-fast and radiation-hard electronic circuits.\citep{Cher2014,Gao2017,Liu2017,Geremew2019,Khitun2018} In modern electronics, metals are primarily used as passive conductors and usually do not play any \emph{active} role. Nanoscale materials with their distinctive size dependent properties provide opportunities to achieve unexplored device functionalities. Ta-based di-chalcogenides, which form layered structures and exhibit charge density wave (CDW),\citep{Wilson1975,Williams1976} are particularly promising in the context. For example, it has been recently shown that 2H-TaSe\tsub2, while showing excellent conductivity properties, also exhibits strong photoluminescence and long photo-carrier lifetime, allowing active metal-based device applications.\citep{Mahajan2019light} CDW is a macroscopic state which is exhibited by materials with reduced dimension, for example, one-dimensional and layered two-dimensional crystals. It is a result of modulation in the electronic charge arising due to a periodic modulation in the crystal lattice. The 1T polymorph of another layered Tantalum di-chalcogenide, namely, TaS\tsub2, exhibits one of the strongest known CDW characteristics enabling temperature dependent distinct conductivity phases.\citep{Wilson1975,Williams1976} Under near-equilibrium heating cycle, 1T-TaS\tsub2 undergoes commensurate (\textit{C}) to triclinic (\textit{T}) phase transition at $223$ K, \textit{T} to nearly-commensurate (\textit{NC}) phase transition at $283$ K, \textit{NC} to incommensurate (\textit{IC}) phase transition at $353$ K and \textit{IC} to normal metal beyond $550$ K.\citep{Manzke1989,Ishiguro1991,Thomson1994} When driven away from equilibrium condition, either by pulsed optical excitation or by an electric field, the crystal exhibits multiple metastable phases with different resistivities.\citep{Stojchevska2014,Yoshida2015,Vaskivskyi2015} In particular, the electrically driven phase transitions in 1T-TaS\tsub2, which result from joule heating, have attracted a lot of attention.\citep{Hollander2015,Liu2016,Yoshida2017,Zheng2017,Zhu2018} This makes 1T-TaS\tsub2 a promising candidate for realizing various functionalities enabling metal-based electronics.

Negative differential resistance (NDR) is an important analog functionality widely used for a variety of applications including high-frequency microwave devices, local oscillators, amplifiers, switching circuits, and multi-valued logic. There have been several attempts of implementation of NDR using two-dimensional layered materials through Esaki diodes \citep{Yan2015,Roy2015,Roy2016,Nourbakhsh2016,Fan2019} and resonant tunneling diodes (RTD) \citep{Britnell2013,Nguyen2014,Lin2015} by exploiting the atomically abrupt junctions - an important requirement to achieve high tunneling efficiency. In both cases, the carriers tunnel through a vertical junction between the valence band and the conduction band (for Esaki diode) or through a tunnel barrier (for RTD). Since both these working principles involve tunneling of carriers through a junction, the peak current density suffers significantly since it is difficult to achieve high tunneling efficiency in two-dimensional layered materials due to (a) relatively large effective mass, particularly in the out-of-plane direction, and (b) misalignment of crystals in vertical heterojunctions resulting in momentum mismatch. For Esaki diodes, a broken gap van der Waals heterojunction has been used recently to mitigate this.\citep{Shim2016} Nonetheless, low peak current density remains a key challenge in the layered material based NDR devices and is undesirable for many applications.

In this work, by exploiting electrically driven phase transitions in 1T-TaS\tsub2, we demonstrate a T-shaped asymmetric 1T-TaS\tsub2/2H-MoS\tsub2 heterojunction based negative differential resistor with a high peak current density ($J_p$) among other layered material based NDR implementations. We are also able to effectively tune the peak current ($I_p$) through both electrical and optical gating. TaS\tsub2 acts as an active material in the device and the working principle relies on the electrically driven phase transition in TaS\tsub2. On the other hand, the semiconducting MoS\tsub2 component in the device adds the gate voltage and light tunability in the obtained $I_p$. The mechanism does not depend on band-to-band tunneling effect and is extremely robust against interface quality, ambiance, and time induced degradation.

\section{Results and discussion}
Figure 1(a) schematically depicts the proposed device structure which is an asymmetric T-junction of multi-layer 1T-TaS\tsub2 and multi-layer 2H-MoS\tsub2. To fabricate the T-junction devices, multi-layer MoS\tsub2 is first transferred on a doped silicon substrate covered by $285$ nm thick thermally grown SiO\tsub2. This is followed by a transfer of the multi-layer TaS\tsub2 flake such that there is a strong asymmetry in length across the two arms of TaS\tsub2. The asymmetry and orientation between the two flakes are controlled by a combination of rotational and translational stages during the transfer process. Ni/Au electrodes are then patterned and deposited using the standard nano-fabrication process. The details of the device fabrication are provided in \textbf{Methods} section. An optical image and the thickness profile of a T-junction (device D\tsub1) after complete fabrication is shown in figure 1(b), where $L\tsub1$ and $L\tsub2$ ($L\tsub1 < L\tsub2$) are respectively the lengths of the top and bottom TaS\tsub2 arms. Figure 1(c)-(d) show the current-voltage characteristics of the heterojunction (T\tsub{1}M) probed between terminals T\tsub{1} and M (with terminal T\tsub{2} kept open) for different values of back-gate voltage ($V_g$) at $290$ K and $180$ K, respectively. The obtained on/off ratio is $\sim 2.8 \times 10^4 $ at $290$ K and $\sim 2.4 \times 10^5 $ at $180$ K with $V_d = -1 V$. Note that with $V_{T_1} > 0$, Ni contact acts as the source and injects electrons into the MoS\tsub2 channel, as explained in the top right inset of figure 1(c). On the other hand, with $V_{T_1} < 0$, it is the TaS\tsub2 that acts as the source and injects electrons into the MoS\tsub2 channel (top left inset). The drive current does not vary significantly for positive and negative $V_{T_1}$, suggesting similar carrier injection efficiency of TaS\tsub2 and Ni contacts to MoS\tsub2.\citep{Mahajan2019} The result emphasizes the possible application of TaS\tsub2 as an efficient van der Waals contact which can maintain the pristine quality of the underneath layered semiconductor. Also, the traces exhibit no hysteresis between the forward and the reverse sweeps at both temperatures, suggesting traps do not play a significant role in these measurements. In analogue to T\tsub{1}M, the current-voltage characteristics of T\tsub{2}M heterojunction (probed between terminals T\tsub{2} and M with terminal T\tsub{1} kept open) at $290$ K and $180$ K are outlined in \textbf{Supporting Figure S1}, which show characteristics similar to the T\tsub{1}M case.

Phase transition induced resistance switching in 1T-TaS\tsub2 has been extensively studied.\citep{Hollander2015,Yoshida2015,Liu2016,Yoshida2017,Zheng2017,Zhu2018} The nearly-commensurate (\textit{NC}) to incommensurate (\textit{IC}) phase transition that normally occurs at 353 K under low electric field condition (quasi-equilibrium) can be electrically driven while operating at room temperature through joule heating. Figure 1(e) shows the current-voltage characteristics of the two-terminal TaS\tsub2 device (by probing terminals T\tsub1 and T\tsub2 of device D\tsub1), delineating the \textit{NC}-\textit{IC} phase transition at $1.8$ V. As we increase the voltage bias across TaS\tsub2 terminals, current-voltage characteristics deviate from linear to non-linear regime beyond $\sim 1$ V. With further increase in the voltage, the current increases abruptly by a factor of $1.56$ which is in accordance with the change in the resistance during \textit{NC}-\textit{IC} phase transition.\citep{Sipos2008,Yu2015,Yoshida2015} This is further evident from the simulation outlined in \textbf{Methods} and results in \textbf{Supporting Figure S2} that the joule heating induced increment in temperature in the channel is sufficient to result in \textit{NC}-\textit{IC} phase transition.
\\
\\
\textbf{Gate tunable NDR and role of asymmetry:} Exploiting the external bias driven phase transition in 1T-TaS\tsub2, we achieve a gate tunable NDR in the 1T-TaS\tsub2/2H-MoS\tsub2 asymmetric T-junction. The equivalent circuit of the $4$-terminal device D\tsub1 embedded with crystal structure of 2H-MoS\tsub2 as well as 1T-TaS\tsub2 in \textit{NC} and \textit{IC} phase is schematically represented in figure 2(a) and 2(b) respectively, wherein $R_1$ and $R_2$ correspond to the TaS\tsub2 resistance in the arms above and below the junction region, and $R_M$ corresponds to the MoS\tsub2 resistance. The fourth terminal is connected to the back gate. The idea is to drive a high electrical field across the TaS\tsub2 terminals T\tsub1 and T\tsub2, thereby inducing the \textit{NC}-\textit{IC} phase transition that in turn reduces the resistance of the TaS\tsub2 branches, and simultaneously monitor the current through the MoS\tsub2 branch ($I_M$). As the voltage $V_{T_1}$ increases and the \textit{NC}-\textit{IC} phase transition occurs, the current through $R_2$ increases abruptly which causes an abrupt reduction in $I_M$, thereby enabling NDR in the MoS\tsub2 branch.

If $n_1$ and $n_2$ are respectively the factors by which the resistance in TaS\tsub2 changes in the upper ($R_1$) and the lower branches ($R_2$) due to the \textit{NC}-\textit{IC} phase transition, the ratio ($\eta$) of $I_M$ before and after the phase transition can be obtained using Kirchoff's law
\beq\label{eq:eta}
\eta = \frac{I_M|_{\textrm{before transition}}}{I_M|_{\textrm{after transition}}} \approx \frac{1+(\frac{n_2}{n_1})\frac{R_1}{R_2}}{1+\frac{R_1}{R_2}}
\eeq
In Equation \ref{eq:eta}, we have used $R_M \gg R_1, R_2$ as observed from figure 1(c) and 1(e). One immediately realizes from the equation that it is important to maintain the asymmetry in the TaS\tsub2 branches to obtain NDR characteristics (that is, $\eta >1$), and a perfectly symmetric structure would result in $\eta=1$ and thus cannot cause NDR.

Figure 2(c) shows the gate voltage dependent NDR characteristics from device D\tsub1 for $V_{T_1}> 0$ (in blue traces) and $V_{T_1}< 0$ (in red traces). The corresponding PVCR values are plotted in figure 2(d). For $V_{T_1}> 0$, the electrons flow from MoS\tsub2 to TaS\tsub2 at the junction without any appreciable energy barrier. Owing to the excellent gate modulation of the MoS\tsub2 channel resistance, we observe a strong gate voltage dependence of the peak current (region A and B), and $I_p$ reaches as high as $4$ $\mu$A $\mu$m$^{-1}$ at $V_g=50$ V from $0.04$ $\mu$A $\mu$m$^{-1}$ at $V_g=-60$ V as shown in figure 2(e) (filled blue spheres). The peak current density of the NDR device can be defined as $J_p = I_p/(W \times L_T)$ where $W$ and $L_T$  denote the width of the channel and transfer length of the MoS\tsub2/TaS\tsub2 contact respectively, as explained in figure 2(a)-(b). We obtain a high peak current density of $4.15 \times 10^4$ nA $\mu$m$^{-2}$ from device D\tsub1 at room temperature which is superior to typical layered material based NDR devices. The device D\tsub1 exhibits a PVCR of about $1.06$. Interestingly, the PVCR shows a very weak dependence on $V_g$ (top-filled blue pentagon in figure 2(d)), which is in good agreement with Equation \ref{eq:eta}, as $\eta$ is independent of $V_g$.

In case of $V_{T_1} < 0$ (the red traces), that is when TaS\tsub2 injects electrons into MoS\tsub2, the electrons need to overcome the TaS\tsub2/MoS\tsub2 Schottky barrier. It has been recently demonstrated that during an \textit{NC}-\textit{IC} phase transition, the change in TaS\tsub2 resistance is accompanied by a suppression in the Schottky barrier height (SBH) at the TaS\tsub2/MoS\tsub2 junction.\citep{Mahajan2019} These two effects compete with each other in the final device characteristics since a suppression in the SBH would result in an abrupt enhancement in the $I_M$, rather than NDR. For $V_g \leq -10$ V (region C), the barrier height suppression plays the dominant role, and thus instead of NDR, we observe an increment in the current $I_M$, as indicated by a PVCR of $< 1$ in figure 2(d). However, with an increase in $V_g$ (region D), the electrons can tunnel through the TaS\tsub2/MoS\tsub2 Schottky barrier due to strong band bending (see figure 2(f)), hence the barrier height suppression does not play a significant role, and the NDR characteristics reappear. The simultaneous modulation of the barrier height at the TaS\tsub2/MoS\tsub2 junction along with the resistivity switching during \textit{NC}-\textit{IC} phase transition provides an additional tunability in the NDR device when $V_{T_1} < 0$.

Now, if we apply the bias at terminal T\tsub2 while connecting $T_1$ and M to a common ground, then the NDR gets completely suppressed for both $V_{T_2}> 0$ and $V_{T_2}< 0$, even for large positive gate voltage (see \textbf{Supporting Figure S3}). This can be easily explained using Equation \ref{eq:eta}, keeping in mind that $R_1$ and $R_2$ are now exchanged. The second term in both numerator and denominator are now small compared to $1$, leading to $\eta \approx 1$, which suppresses any NDR.

Note that the \textit{NC}-\textit{IC} phase transition is accompanied by hysteresis (see, for example, figure 1(e)) which manifests as a hysteresis window in the NDR characteristics as well. The hysteresis window can be reduced by introducing fast heat dissipation channels in the device. We fabricate a second device (D\tsub2) where the MoS\tsub2 layer underneath is a wide one, facilitating improved heat dissipation, as explained in \textbf{Supporting Figure S4}. The external bias dependent current-voltage characteristics of the TaS\tsub2 sample exhibit negligible hysteresis. This manifests in the reduced hysteresis window when the device is operated in the NDR mode, both for $V_{T_1}> 0$ and $V_{T_1}< 0$. Note that the two TaS\tsub2 branches are more symmetric in D\tsub2 than in D\tsub1, which results in a slightly reduced value of the PVCR, further verifying the operation principle described earlier.
\\
\\
\textbf{Light tunable NDR:} Before demonstrating the light tunable NDR effect, we first show the efficient phototransistor effect by light gating achieved in the heterojunction. The configuration is shown in figure 3a. Figure 3b shows the photo-induced current-voltage characteristics of the heterojunction T\tsub1M (probed between terminal T\tsub1 and M with terminal T\tsub2 kept open) at $V_g = -50$ V (See \textbf{Supporting Figure S5} for characteristics at other gate voltage values). When excited with a laser of wavelength $532$ nm and a spot size encompassing the entire device, the photocurrent across the device can be effectively modulated with the excitation power. In particular, when the device is turned off by applying $V_g=-60$ V, with an excitation power ($P_{op}$) of $45.4\ nW$, the device current switches by a factor of $1.65 \times 10^2$ for $V_{T_1}=1$ V, and $2.52 \times 10^3$ for $V_{T_1}=-1$ V from the dark condition. The responsivity of the device is extracted as $R = \frac{I_{light}-I_{dark}}{P_{op}}$ where $I_{light}$ and $I_{dark}$ are the current with and without light, respectively. The calculated $R$ is plotted as a function of incident optical power for $V_{T_1} = -1$ V in figure 3(c). $R$ increases with a reduction in the excitation power, and reaches a value as high as $2.31 \times 10^{4}\ AW^{-1}$ at an excitation power of $0.22\ nW$.

We now exploit this photo-induced gating effect in the 4-terminal measurement to demonstrate light tunable NDR. As before, during this measurement, bias is applied in the terminal T\tsub1 for different values of the back gate voltage, keeping terminals T\tsub2 and M grounded. The results are shown in figure 3(d)-(f) for $V_g=-50$, $-20$, and $10$ V with $V_{T_1} > 0$ V. We observe that the NDR characteristics can be effectively tuned by the excitation power owing to a photogating induced modulation in the resistance of the MoS\tsub2 branch with optical power. Given the low power of the incident optical excitation, we do not expect any modulation of TaS\tsub2 conductivity with light. The simultaneous control of $I_p$ by $V_g$ and $P_{op}$ is depicted in the color plot in figure 3g, which suggests that the degree of photo-tunability of the NDR is a strong function of the applied back gate voltage. Even when $V_g=-50$ V, with photo-excitation, we are able to achieve a peak current which is in excess of $1\ \mu$A$\ \mu$m$^{-1}$, with a PVCR of $\sim 1.065$. Such a light and gate voltage tunable NDR characteristics are promising for light-gated optoelectronic applications. For $V_{T_1} < 0$ V, where TaS\tsub2 injects electrons to MoS\tsub2 over an SBH, we obtain an optical power-dependent enhancement in the current rather than NDR effect at the phase transition, as explained in \textbf{Supporting Figure S6}. This is in agreement with the phase transition induced TaS\tsub2/MoS\tsub2 barrier suppression effect discussed earlier.
\\
\\
\textbf{Enhancing the PVCR:} An enhanced switching ratio during the TaS\tsub2 phase transition is desirable to further improve the PVCR in the proposed T-junction. In order to do so, we exploit electrically driven phase transition in TaS\tsub2 at low temperature. \citep{Yoshida2015, Zhu2018, Liu2016} In addition to the equilibrium states, 1T-TaS\tsub2 exhibits several metastable states that can be accessed by electric field \citep{Yoshida2015} and optical pulse excitation.\citep{Stojchevska2014, Vaskivskyi2015} Figure 4(a) shows the current-voltage characteristics when bias is applied across the two terminals of TaS\tsub2 (T\tsub1 and T\tsub2) at $160$ K, $170$ K and $180$ K. The base sample temperature is intentionally kept below the \textit{C}-\textit{T} transition temperature to electrically drive the system into metastable state and eventually drive to \textit{IC} state, which causes an abrupt increase in the current by a factor of $\sim 2.23$.

In the corresponding four-terminal measurement at $180$ K, $I_M$ is plotted as a function $V_{T_1}$ in figure 4(b). For $V_{T_1}>0$, the peak current shows a strong modulation by the gate voltage. The obtained PVCR is plotted in figure 4(c) as a function of $V_g$, and is found to show a weak $V_g$ dependence as expected from Equation \ref{eq:eta}, and saturates around $1.2$, which is an improvement over the value obtained at room temperature. In contrast to figure 2(d), the peak-to-valley transition at low temperature depicts multiple transitions (see top right inset of figure 4b), which results from the existence of the multiple metastable states in 1T-TaS\tsub2 under non-equilibrium.

For $V_{T_1} < 0$, similar to figure 2(d), we obtain two distinct regions of operation. When $V_g \leq -10$ V, we obtain a strong enhancement in $I_M$ due to the phase transition, and enhancement factor increases for larger negative $V_g$. In the left inset of figure 4(b), we show $I_M$ in log scale, delineating the abrupt increment in the current, which is enhanced compared with what we obtained at room temperature in figure 2(c). The stronger enhancement in $I_M$ and its gate voltage dependence is a clear indication of a strong suppression of the SBH at the TaS\tsub2-MoS\tsub2 interface during the meta-stable state to \textit{IC} phase transition. On the other hand, when $V_g \geq 10$ V, the electrons tunnel through the Schottky barrier due to strong band bending and the barrier suppression effect plays a less significant role. Under this condition, the NDR effect reappears due to the phase transition induced resistance switching of TaS\tsub2.

There is also a third regime of operation for $V_{T_1} < 0$ around $V_g=0$ V, where we obtain a high PVCR of $1.59$ (figure 4(c)) which cannot be explained just by the phase transition induced current division effect. This region of operation is completely absent for $V_{T_1} > 0$. We also note a hysteresis associated with the current-voltage characteristics for negative $V_{T_1}$. This suggests the role of carrier trapping at the TaS\tsub2/MoS\tsub2 junction in the increased PVCR. During the TaS\tsub2 phase transition, the local temperature at the junction increases abruptly due to a sudden increase in the current $I_{T_1}$. This is supported by the simulation predicted temperature at the junction during the voltage sweep, as shown in figure 4(d). This results in activated local trapping of carriers, which in turn causes a screening of the gate field, reducing the magnitude of $I_M$. Note that with an increment in the gate voltage to higher positive values, the carriers tunnel through this barrier, and thus such trapping effects do not play a significant role.

The above discussion on the enhancement of PVCR using CDW phase transition at low temperature provides key insights about the device functionality, and has important implication on several niche applications (for example, space electronics) where both low temperature operation and radiation hardness are necessary. However, the improved performance at low temperature may have limited impact on applications requiring room temperature operation. In order to increase the PVCR at room temperature, the upper TaS\tsub2 branch ($R_1$) of the T-network in Figure \ref{fig2}a-b can be replaced by a constant resistance where no such phase transition occurs. This forces $n_1$ in Equation \ref{eq:eta} to unity, which thus enhances $\eta$ and hence, the PVCR. In such a design, the PVCR will have an upper limit of $n_2$ with a suitable design ensuring $R_1>>R_2$.

While the additional terminals in the proposed multi-terminal device provide gate control and access to more state variables, from an implementation point of view, it would require some changes in the conventional NDR-based circuit design. For example, to implement an oscillator, where the positive resistance of a tank circuit is compensated by the negative resistance, the top (T\tsub1) and right (M) ports (see Figure \ref{fig2}a-b) can be directly used, grounding terminal T\tsub2.

The heterojunctions studied here are robust against ambience induced degradation and exhibit good repeatability of the data measured over more than a month (see \textbf{Supporting Figure S7}).In Table \ref{table:1} we benchmark the performance of the devices reported here against other NDR devices based on both layered materials and bulk semiconductors. The peak current density of the proposed device exceeds the numbers reported from layered materials based tunneling devices \citep{Yan2015,Roy2015,Roy2016,Nourbakhsh2016,Fan2019,Britnell2013,Nguyen2014,Lin2015,Shim2016}, and also compares well with several bulk technologies \citep{Oehme2010,Oehme2009,Oehme2010Ge,Park2009,Duschl2000,Golka2006}. With further optimization, the proposed device is promising to be comparable with SiGe and III-V semiconductor based complex multi-heterojunctions with highest reported peak current densities \citep{Bayram2010,Chung2006,Pawlik2012}.

Active metal based electronics, such as the current design, is expected to provide radiation hardness. In a conventional tunnel diode, a major cause for performance degradation due to exposure to radiation is an increase in the excess current, and in turn a degradation in the PVCR \citep{Soliman1994}. This is primarily due to a degradation in the tunnel junction interface due to defect generation, resulting in an increase in the trap assisted tunneling \citep{Majumdar2014}. The proposed device is expected to have an advantage in this aspect since it avoids any tunnel junction. Equation \ref{eq:eta} suggests that the PVCR in the device only depends on the resistance components (and switching ratios) of the TaS\tsub2 branches, and independent of the semiconducting MoS\tsub2 component, and thus the PVCR is inherently radiation hard. Finally, one could, in principle, replace the semiconducting MoS\tsub2 component by a metallic branch of appropriate resistance to make it more radiation hard. The PVCR can still be maintained based on Equation \ref{eq:eta}, though the optical control or the electrical gate control offered by the semiconducting component will be lost.

\section{Conclusion:}
In conclusion, we demonstrated a different way of achieving negative differential resistance by simultaneously exploiting the resistivity and Schottky barrier height switching using an electrically driven phase transition arising from charge density wave in a 1T-TaS\tsub2/2H-MoS\tsub2 heterojunction. The negative differential resistance can be effectively tuned by external stimuli including photo-excitation and a back gate voltage. The design of the device distinguishes itself from other tunneling based NDR implementations by using majority carriers and not requiring any stringent interface quality - thus achieving a high peak current density. The proposed device also exhibits excellent air stability and repeatability. The results are promising towards the implementation of various functionalities employing robust and tunable negative differential resistance, and mark a step towards 1T-TaS\tsub2 based metal electronics.

\section{Methods}
\textbf{T-junction device fabrication and characterization:} The bulk crystals of 1T-TaS\tsub2 and 2H-MoS\tsub2 are procured from the 2D Semiconductors. 1T-TaS\tsub2/2H-MoS\tsub2 T-junction is fabricated in two steps: first, the thin MoS\tsub2 flakes are mechanically exfoliated from scotch tape onto a heavily doped Si substrate coated with $285$ nm thick SiO\tsub2 using polydimethylsiloxane (PDMS). Second, the TaS\tsub2 flakes are transferred through PDMS attached to a glass slide that is controlled by an additional manipulator. The alignment of the TaS\tsub2 flake with respect to the MoS\tsub2 flake is controlled by the rotational stage in order to form the T-junction. The whole exfoliation and transfer processes are done at room temperature, with no additional cleaning step. The substrate is then spin coated with a high contrast positive resist $ - $ polymethyl methacrylate (PMMA) $950$ C$3$ and softly baked for $2 $ minutes at $180$ $^\circ$C for pattern writing by means of electron beam lithography with an electron beam dose of $200$ $\mu$C cm$^{-2}$, an electron beam current of $300$ pA, and an acceleration voltage of $20$ KV. The pattern development is carried out in $1:3$ MIBK/IPA developer solution followed by IPA wash and blow drying in N\tsub2. Metal contacts are formed by blanket deposition of $ 10 $ nm Ni / $ 50 $ nm Au using a DC magnetron sputter coating system in presence of Ar plasma at $ 6.5\times 10^{-3} $ Torr. Excess metal lift-off is carried out by immersing the substrate in acetone for $ 15-30 $ minutes followed by IPA wash for $ 30 $ seconds and blow drying in N\tsub2. Buffered HF solution is used to etch the back oxide from the substrate and highly conducting silver paste is used for the back gate contact.
\\
The electrical measurements are performed in a probe station with a base vacuum level of about $3.45 \times 10^{-3}$ Torr at room temperature. The low temperature measurements are performed with liquid N\tsub2 supply at a base vacuum of $6.45 \times 10^{-6}$ Torr. Measurements under light illumination are performed using a $532$ nm laser, with the spot size being larger than the device area. The optical power is calibrated for the area factor and the loss due to the transmission through the optical window of the probe station.
\\\\
\textbf{Thermal model Simulation:} The temperature variation along the TaS\tsub2 channel resulting from joule heating is obtained by solving the one-dimensional heat equation described below using finite element method:\citep{Grosse2011,Zhu2018,Geremew2019ACS}
\begin{equation}\label{eq:1}
A\frac{d}{dx}\left(K\frac{dT}{dx}\right) + P = g(T-T_{base})
\end{equation}
where
\begin{conditions}
	P 			&  Heat generation rate per unit channel length.\\
	A			&  Cross-sectional area of the device.\\
	K 			&  Thermal conductivity of 1T-TaS\tsub2.\\
	T_{base}	&  Base temperature.\\
	g 			& Thermal conductance of the substrate per unit channel length.\\
\end{conditions}
This results in a non-uniform rise in temperature along the channel. We incorporate the temperature dependence of the thermal conductivity ($K$) \citep{Regueiro1985} in the model. At high bias, the temperature increases abruptly in the middle of the channel ($x = 4.68\ \mu m$) during the phase transition as shown in figure S2(a) and (c) for $298$ K and $180$ K respectively. The simulated temperature profile across the TaS\tsub2 channel shows strong non-uniformity for a base temperature of $298$ K and $180$ K at different biasing conditions (see figure (b) and (d) of \textbf{Supporting Figure S2}). The temperature drops near the ends of the channel due to the heat sink nature of Ni/Au contact. However, the temperature in the middle of the channel increases with the external bias which confirms the non-uniformity of the Joule heating induced temperature change. Owing to the non-equilibrium heat generation in the TaS\tsub2 channel, the phase transitions are achieved through multiple metastable states.

\begin{acknowledgement}
	
	K.M. acknowledges the grant from Indian Space Research Organization (ISRO), grant under Ramanujan Fellowship, the Early Career Award, and Nano Mission from the Department of Science and Technology (DST), and support from MHRD, MeitY and DST Nano Mission through NNetRA.
	
\end{acknowledgement}

\section{Associated Content}
	The authors declare no competing financial or non-financial Interests.

\begin{suppinfo}	
Supporting information available on the following:
Electrical Characterization of TaS\tsub2/MoS\tsub2 junction; Simulated temperature variation along TaS\tsub2 channel; Characteristics for T\tsub2 biasing in device D\tsub1; Gate tunable NDR with reduced hysteresis in 1T-TaS\tsub2/2H-MoS\tsub2 asymmetric T-junction device (D\tsub2); Light response of heterojunction T\tsub1M of device D\tsub1; Light controlled NDR in TaS\tsub2/MoS\tsub2 asymmetric T-junction (D\tsub1); Stability of TaS\tsub2/MoS\tsub2 asymmetric T-junction (D\tsub1).
\end{suppinfo}

\newpage
\bibliography{NDR_1}

\newpage
\begin{figure}[!hbt]
	\centering
	\includegraphics[scale=0.5]{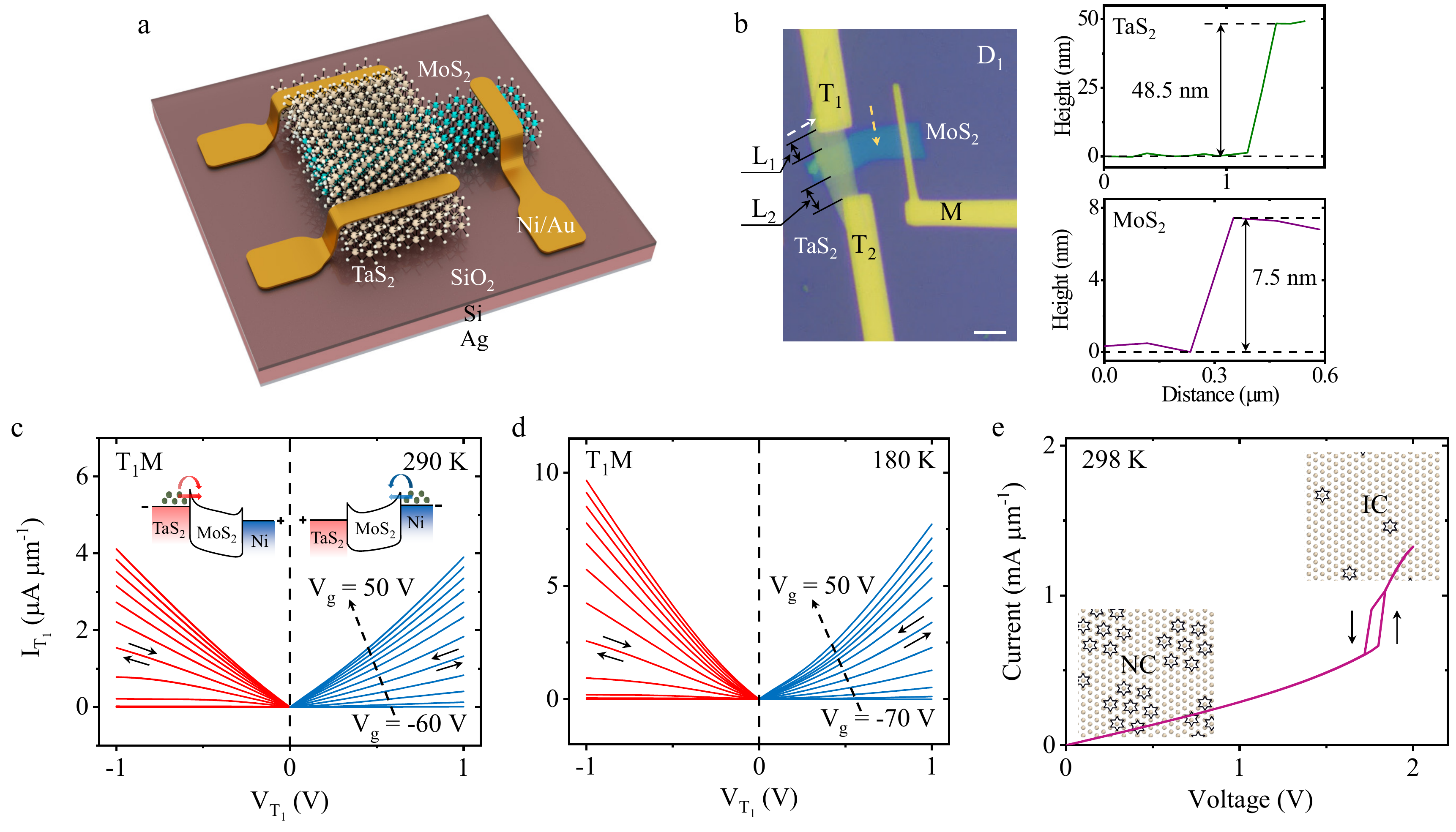}
	\caption{\textbf{Electrical Characterization of 1T-TaS\tsub2/2H-MoS\tsub2 asymmetric T-junction.} (a) Schematic representation of the T-Junction. (b) Left panel: Optical image of the fabricated asymmetric T-junction device (Scale bar: $5\ \mu m$). $L_1$ and $L_2$ denote the length of the TaS\tsub2 arm above and below the junction, respectively. Right panel: Thickness profile of the TaS\tsub2 (in top) and MoS\tsub2 (in bottom) flakes along the white and yellow dashed arrows indicated in the optical image. (c)-(d) Current-voltage characteristics of TaS\tsub2/MoS\tsub2 junction (T\tsub1M) probed between terminals T\tsub1 and M (with terminal T\tsub2 open) as a function of $V_g$ varying from $-60\ V$ to $50\ V$ in steps of $10\ V$ at $290$ K [in (c)] and at $180$ K [in (d)]. Inset of (c): For $V_{T_1} > 0$, Ni acts as a source of electrons, while for $V_{T_1} < 0$, TaS\tsub2 sources electrons into the MoS\tsub2 channel. Forward and reverse sweeps are indicated by black arrows. (e) Current-voltage characteristics of 2-probe 1T-TaS\tsub2 device (probed between terminals T\tsub1 and T\tsub2) depicting NC-IC phase transition at 298 K. Inset: Crystal structure of 1T-TaS\tsub2 in NC phase (bottom left) and IC phase (top right).}
	\label{fig1}
\end{figure}

\newpage
\begin{figure}[!hbt]
	\centering
	\includegraphics[scale=0.5]{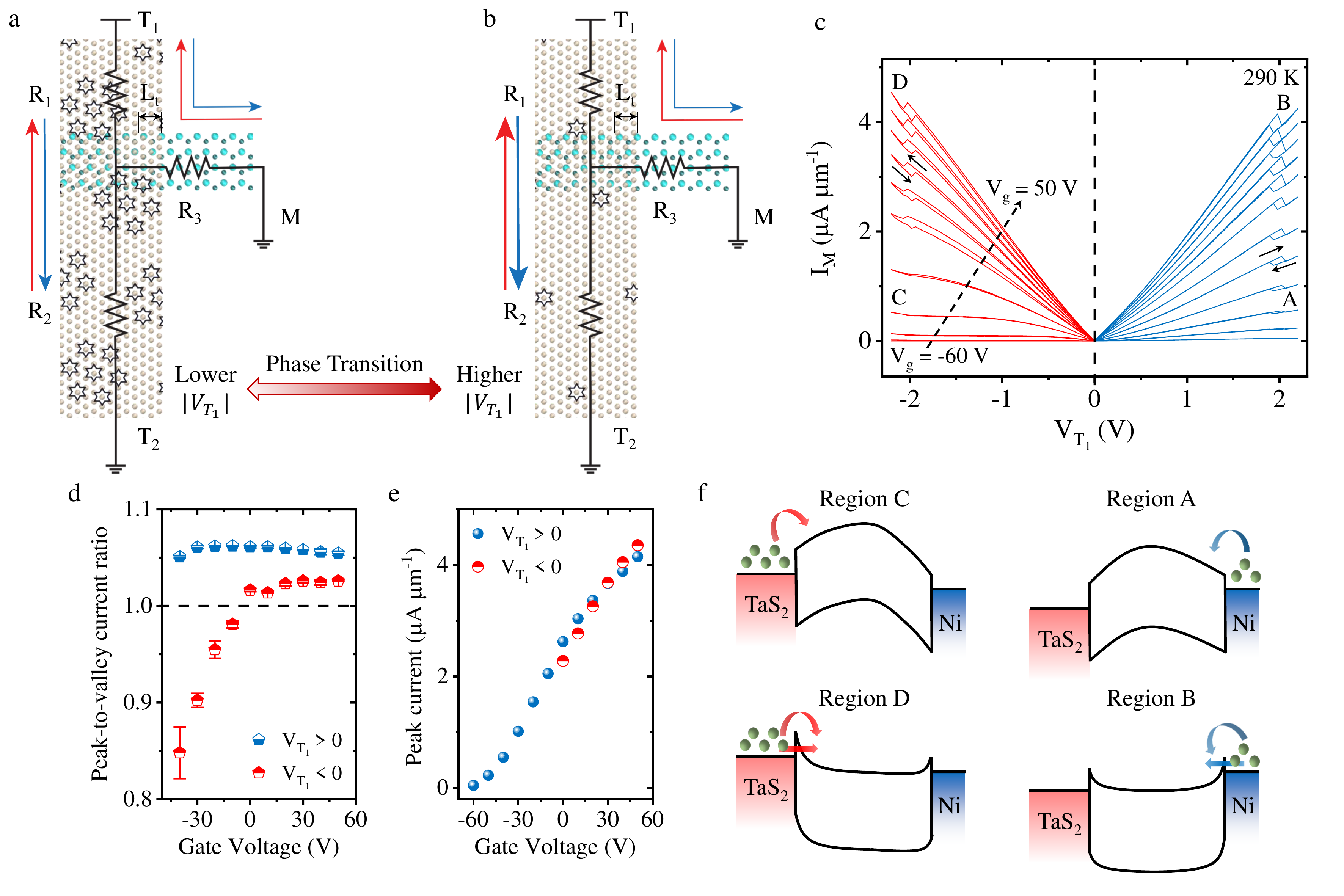}
	\caption{\textbf{Gate tunable Negative Differential Resistance (NDR).} (a)-(b) Equivalent circuit diagram of the device D\tsub1 embedded with crystal structure of 1T-TaS\tsub2 in NC-phase [in (a)] and IC-phase [in (b)]. Blue and red arrows indicate the direction of current flow when $V_{T_1} > 0$ and $V_{T_1} < 0$, respectively. (c) Current through the MoS\tsub2 branch ($I_M$) \textit{versus} $V_{T_1}$ as a function of $V_g$ varying from $-60\ V$ to $50\ V$ in steps of $10\ V$ at $290$ K depicting NDR for both $V_{T_1} > 0$ and $V_{T_1} < 0$. Forward and reverse sweeps are indicated by black arrows. (d) Peak-to-valley current ratio (PVCR) as the function of $V_g$ for both $V_{T_1} > 0$ (bottom-filled blue pentagon) and $V_{T_1} < 0$ (top-filled red pentagon) at $290$ K. (e) Peak current ($I_p$) \textit{versus} $V_g$ for both $V_{T_1} > 0$ (blue spheres) and $V_{T_1} < 0$ (half-filled red circles) at $290$ K. (f) Band diagram corresponding to different regions of operation, namely, A, B, C and D as indicated in (c).}
	\label{fig2}
\end{figure}

\newpage
\begin{figure}[!hbt]
	\centering
	\includegraphics[scale=0.45]{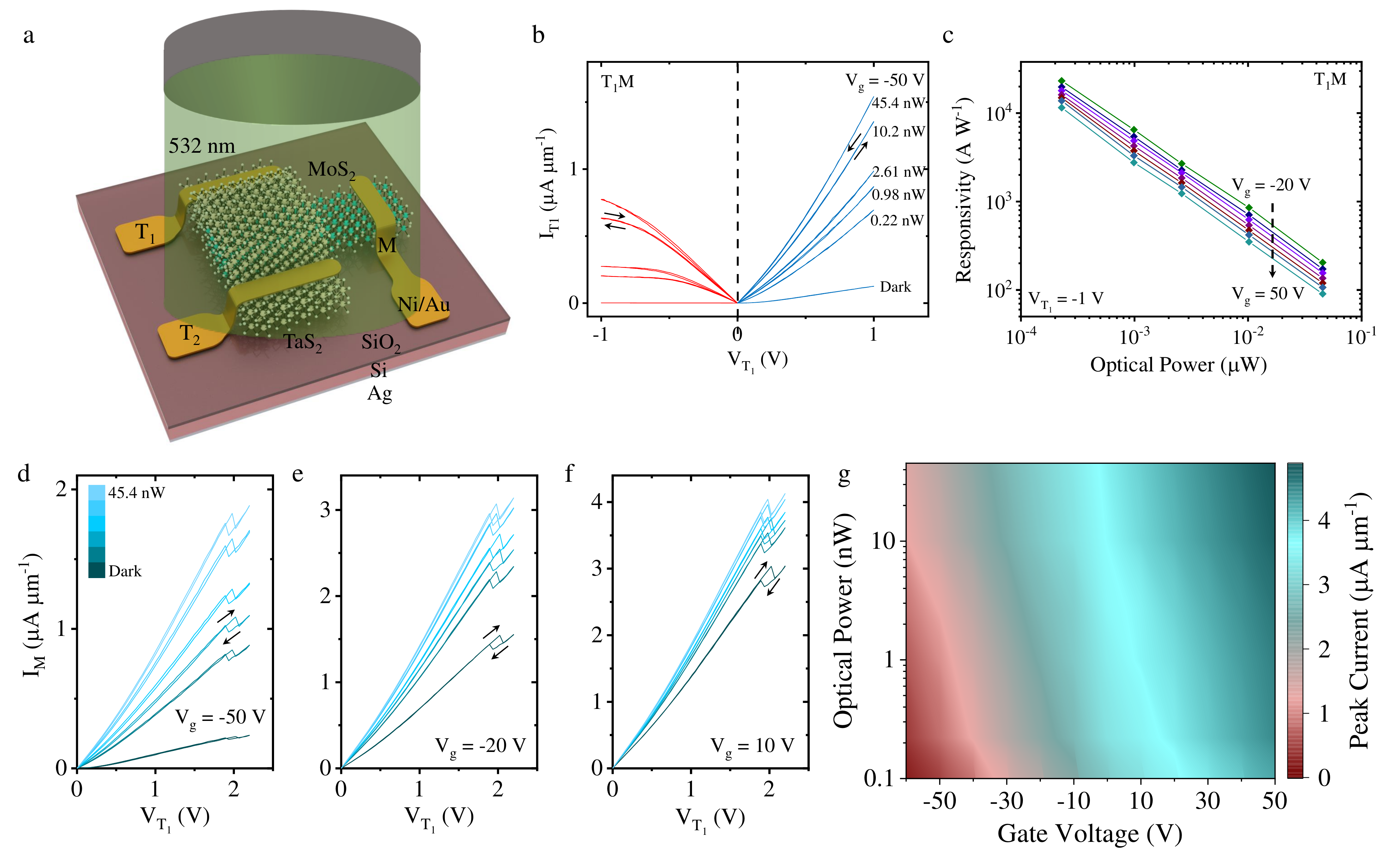}
	\caption{\textbf{Light tunable NDR.} (a) Schematic diagram of the device under photo-excitation. (b) Current-voltage characteristics of heterojunction T\tsub1M (with T\tsub2 open) at $V_g = -50 V$  as a function of 532 nm laser excitation power, showing strong photogating. (c) Responsivity \textit{versus} excitation power at different $V_g$ values for $V_{T_1} = -1\ V$. (d)-(f) $I_M$ \textit{versus} $V_{T_1}$ as a function of excitation power at $V_g = -50\ V$ [in (d)], $V_g = -20\ V$ [in (e)] and $V_g = 10\ V$ [in (f)]. Forward and reverse sweeps are indicated by black arrows. (g) Color plot of peak current in gate voltage-excitation power control space.}
	\label{fig3}
\end{figure}

\newpage
\begin{figure}[!hbt]
	\centering
	\includegraphics[scale=0.5]{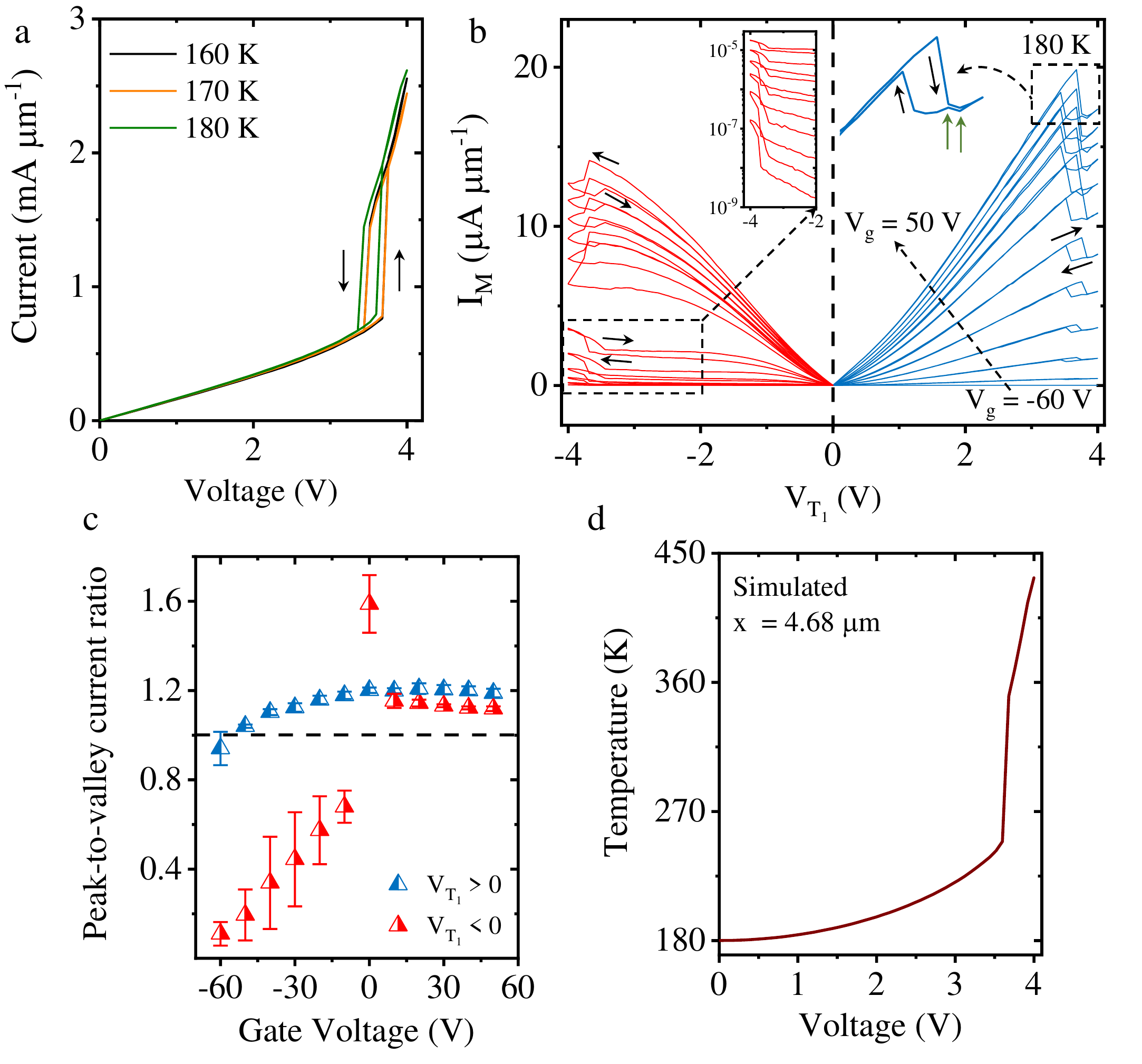}
	\caption{\textbf{Enhancing PVCR at low temperature.} (a) Current-voltage characteristics of 2-probe TaS\tsub2 (probed between terminals T\tsub1 and T\tsub2) at $160$ K (black trace), $170$ K (orange trace) and $180$ K (olive trace). (b) $I_M$ \textit{versus} $V_{T_1}$ as a function of $V_g$ varying from $-60\ V$ to $50\ V$ in steps of $10\ V$ at $180$ K. Left Inset:  $I_M$ \textit{versus} $V_{T_1}$ for $V_{T_1} < 0$ and $V_ g < 0$ in logarithmic scale, showing sharp enhancement in current during phase transition due to SBH suppression effect. Right Inset: Zoomed-in view of NDR region for MoS\tsub2 injection at $V_g = 50\ V$ clearly indicating multi-state induced NDR wherein different transitions are indicated by green arrows. Forward and reverse sweeps are indicated by black arrows. (c) PVCR \textit{versus} $V_g$ for $V_{T_1} > 0$ (left-filled blue triangles) and $V_{T_1} < 0$ (right-filled red triangles) injection at $180$ K. (d) Simulated temperature variation in the middle of TaS\tsub2 channel ($x = 4.68\ \mu m$) as a function of the voltage bias applied. }
	\label{fig4}
\end{figure}

\newpage
\begin{table}[!hbt]
	\centering
	\caption{Performance benchmarking with existing literature}
	\label{table:1}
	\begin{adjustbox}{width=\columnwidth,center}
		\begin{tabular}{|c|c|c|c|c|c|c|c|}
			\hline
			\textbf{Stack} &\textbf{Mechanism} & \textbf{T} & \boldmath{$I_P$} & \boldmath{$J_P$} & \textbf{PVCR} & \textbf{Gate} & \textbf{Light} \\ [0.5ex]
			& & \textbf{(K)} & \textbf{(nA)} & \textbf{(\boldmath{$nA \mu m^{-2}$})} & & \textbf{Control} & \textbf{Control}\\
			&&&&&& \textbf{of} & \textbf{of}  \\ [0.5ex]
			&&&&&& \textbf{NDR} & \textbf{NDR}  \\ [0.5ex]
			\hline
			\multirow{5}{*}{\textbf{MoS\tsub2/TaS\tsub2 (This work)}} & \multirow{5}{*}{\textbf{CDW}} & $300$ & $2.07 \times 10^4$ & $4.15 \times 10^4$ & $1.06$  & \multirow{5}{*}{\textbf{Yes}} & \multirow{5}{*}{\textbf{Yes}}  \\
			\cline{3-6}
			&&  \multirow{4}{*}{$180$} & \multirow{2}{*}{$9.92 \times 10^4$} & \multirow{2}{*}{$1.98 \times 10^5$} & $1.2$ && \\
			&&&&&  $@V_g = 50V$ &&\\
			\cline{4-6}
			&&& \multirow{2}{*}{$4.8 \times 10^4$} & \multirow{2}{*}{$9.61 \times 10^4$} & $1.59$ &&\\
			&&&&& $@V_g = 0V$ &&\\
			\hline
			\multirow{2}{*}{BP/SnSe\tsub2{\citep{Yan2015}}} & \multirow{2}{*}{BTBT\tsup{\emph{a}}} & $300$ & $145$ & $1.45$ & $1.8$ & \multirow{2}{*}{Yes} & \multirow{2}{*}{No}  \\
			\cline{3-6}
			&& $80$ & $175$ & $1.75$ & $2.8$ & &   \\
			\hline
			DG MoS\tsub2/WSe\tsub2 {\citep{Roy2015}} & BTBT & $77$ & $1.26$ & $0.45$ & $1.68$ & Yes & No   \\
			\hline
			WSe\tsub2/SnSe\tsub2 {\citep{Roy2016}} & BTBT & $77$ & $0.27$ & $-$ & $1.03$ & No & No   \\
			\hline
			MoS\tsub2/WSe\tsub2 {\citep{Nourbakhsh2016}} & BTBT & $300$ & $0.24$ & $-$ & $1.55$ & Yes & No  \\
			\hline
			Gr/hBN/WSe\tsub2/SnSe\tsub2 {\citep{Fan2019}} & BTBT & $300$ & $3 \times 10^4$ & $1.46 \times 10^3$ & $4$ & Yes & No     \\
			\hline
			\multirow{2}{*}{hBN/Gr/hBN/Gr{\citep{Britnell2013}}}  & \multirow{2}{*}{RTD\tsup{\emph{b}}} & $300$ & $75$ & $125$ & $1.3$ & \multirow{2}{*}{Yes} & \multirow{2}{*}{No}  \\
			\cline{3-6}
			& & $7$ & $110$ & $183$ & $3.6$ & &   \\
			\hline
			MoS\tsub2 {\citep{Nguyen2014}} & RTD & $10$ & $2.1$ & $-$ & $1.1$ & Yes & No   \\
			\hline
			Gr/WSe\tsub2/MoS\tsub2 {\citep{Lin2015}} & RTD & $300$ & $0.05$ & $-$ & $1.9$ & No & No   \\
			\hline
			Gr/MoSe\tsub2/WSe\tsub2 {\citep{Lin2015}} & RTD & $300$& $0.08$ & $-$ & $2.2$ & No & No  \\
			\hline
			\multirow{2}{*}{BP/ReS\tsub2 {\citep{Shim2016}}} & \multirow{2}{*}{BTBT} & $300$ & $2.7$ & $-$ & $4.2$ & \multirow{2}{*}{Yes} & \multirow{2}{*}{No}  \\
			\cline{3-6}
			&     & $180$ & $3$ & $-$ & $6.9$ & &   \\
			\hline
			Si {\citep{Oehme2010}} & BTBT &	$300$ & $-$ & $4.25 \times 10^1$ & $5.05$ & No & No   \\
			\hline
            Si {\citep{Oehme2009}} & BTBT & $300$ & $-$ & $6.6 \times 10^4$ & $2.5$ & No & No   \\
			\hline
			Ge {\citep{Oehme2010Ge}} & BTBT & $300$ & $-$ & $5.7$ & $1.6$ & No & No   \\
			\hline
			Si/SiGe	{\citep{Park2009}} & RITD\tsup{\emph{c}} & $300$ &	$1.78 \times 10^6$ & $1 \times 10^3$ & $1.85$ & No & No   \\
			\hline
			Si/SiGe/Si {\citep{Duschl2000}} & BTBT & $300$ & $-$ & $3 \times 10^5$ & $4.8$ & No & No   \\
			\hline
			GaN/AlGaN {\citep{Golka2006}} & RTD & $300$ & $-$ & $1 \times 10^5$ & $2$ & No & No   \\
			\hline
			AlN/GaN {\citep{Bayram2010}} & RTD & $300$ & $2.32 \times 10^7$ & $1.2 \times 10^6$ & $1.2$	& No & No   \\
			\hline			
            Si/SiGe	{\citep{Chung2006}}	& RITD & $300$ & $-$ & $2.18 \times 10^6$ &	$1.47$ & No & No   \\
			\hline
            InAs/GaSb {\citep{Pawlik2012}}	& BTBT & $300$ & $-$ & $2.2 \times 10^7$ & $3.9$ & No & No   \\
			\hline
		\end{tabular}
	\end{adjustbox}
	\tsup{\emph{a}} Band-to-band tunneling;
	\tsup{\emph{b}} Resonant Tunneling;
	\tsup{\emph{c}} Resonant Interband Tunneling.
\end{table}


\section*{Supplementary material}

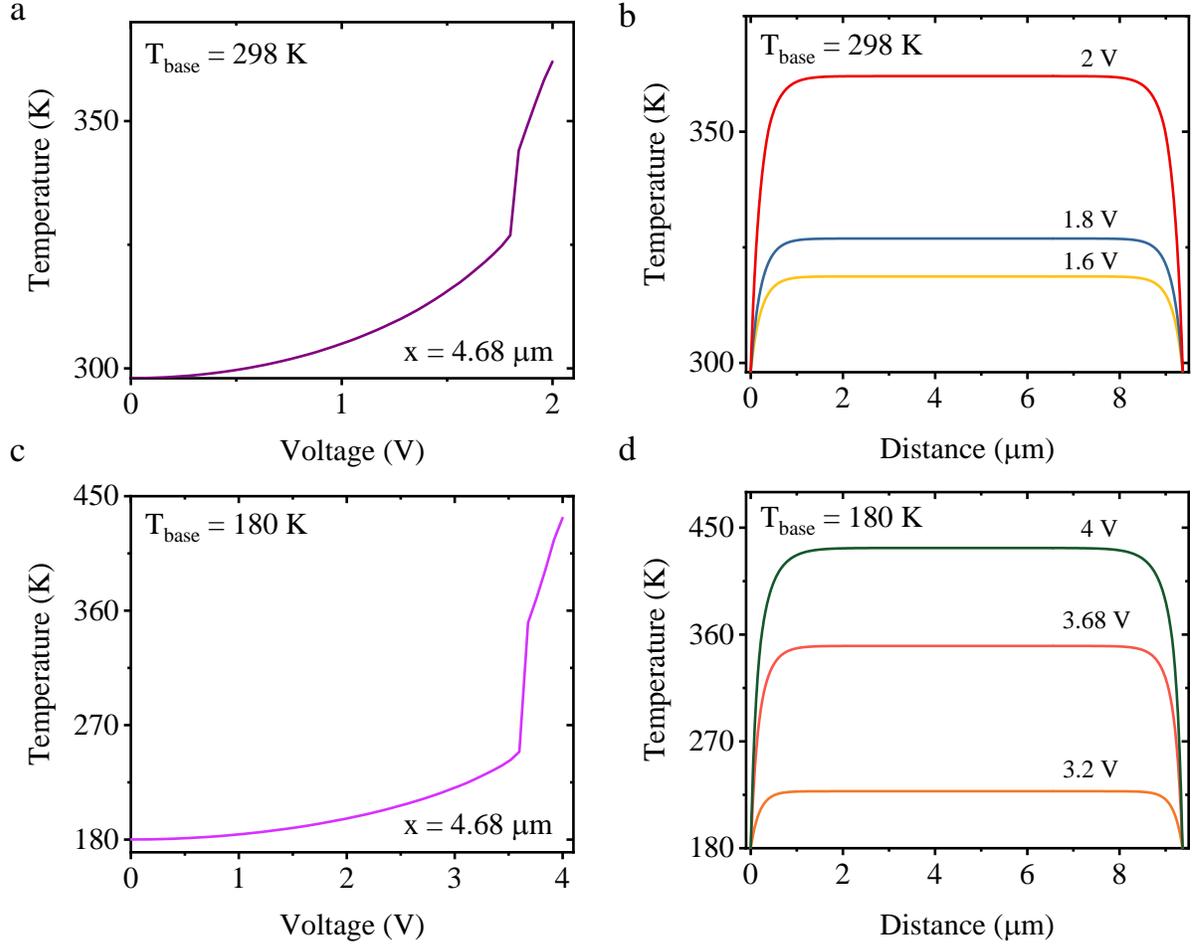

Figure S2: **Simulated temperature variation along TaS$_2$ channel.** (a)-(c) Simulated temperature profile at the middle of the TaS$_2$ channel as the function of voltage for base temperature ($T_{base}$) of 298 K [in (a)] and 180 K [in (c)]. (b) Simulated temperature variation along the TaS$_2$ channel at bias voltage of 1.6 V (before transition), 1.8 V (at transition) and 2 V (after transition) for $T_{base}$ of 298 K. (d) Simulated temperature variation along the TaS$_2$ channel at bias voltage of 3.2 V (before transition), 3.68 V (at transition) and 4 V (after transition) for $T_{base}$ of 180 K.

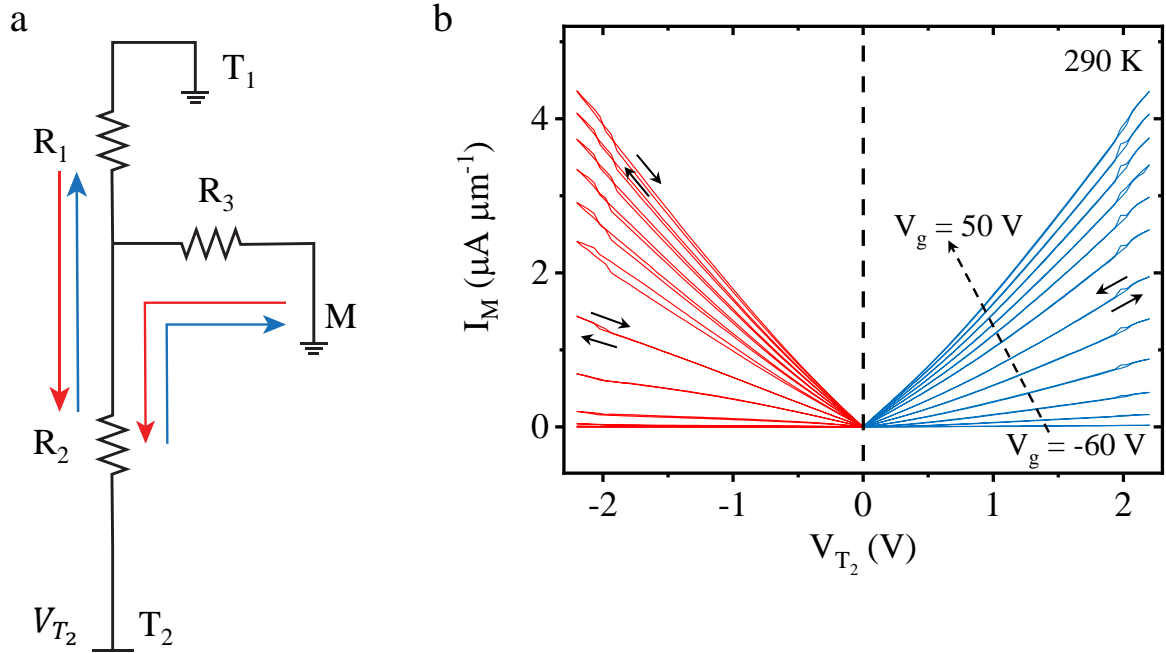

Figure S3: **Characteristics for $T_2$ biasing in device $D_1$.** (a) Equivalent circuit diagram of the 4-terminal device $D_1$ when bias is applied at terminal $T_2$ whereas terminal $T_1$ and M are grounded. (b) $I_M$ *versus* $V_{T_2}$ as a function of back gate voltage ($V_g$) varied from $-60\ V$ to $50\ V$ in steps of 10 V at 290 K depicting a small increment in current for both $V_{T_1} > 0$ (blue traces) and $V_{T_1} < 0$ (red traces). Forward and reverse sweeps are indicated by black arrows. No NDR is observed for this biasing condition, in agreement with Equation 1 in the main text.

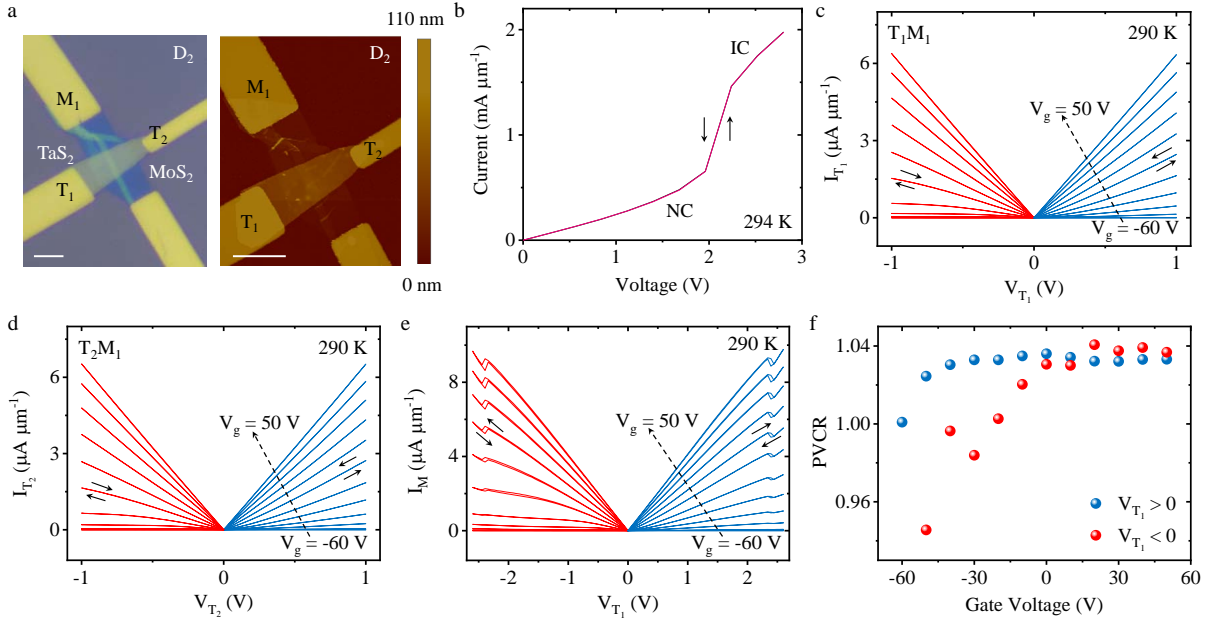

**Figure S4: Gate tunable NDR with reduced hysteresis in 1T-TaS$_2$/2H-MoS$_2$ asymmetric T-junction device (D$_2$).** (a) Optical and AFM image of the fabricated device D$_2$ in left and right panel respectively. Scale bar: 5 $\mu m$. (b) I-V characteristics of 2-probe 1T-TaS$_2$ device D$_2$ (probed between terminals T$_1$ and T$_2$) under high field condition depicting NC-IC phase transition. (c) I-V characteristics of TaS$_2$/MoS$_2$ junction (T$_1$M$_1$) probed between terminals T$_1$ and M$_1$ (with terminal T$_2$ open) as a function of back gate voltage ($V_g$) varied from $-60\ V$ to $50\ V$ in steps of $10\ V$ at 290 K. (d) I-V characteristics of T$_2$M$_1$ junction probed between terminals T$_2$ and M$_1$ (with terminal T$_1$ open) for different $V_g$ at 290 K. (e) $I_M$ versus $V_{T_1}$ as the function of back gate voltage ($V_g$) at 290 K depicting NDR for both $V_{T_1} > 0$ (blue traces) and $V_{T_1} < 0$ (red traces). Forward and reverse sweeps are indicated by black arrows. Reduced hysteresis in the observed NDR is inherited by phase switching curve in (b). (f) Peak-to-valley current ratio (PVCR) as a function of gate voltage ($V_g$) for both $V_{T_1} > 0$ (blue spheres) and $V_{T_1} < 0$ (red spheres).

5

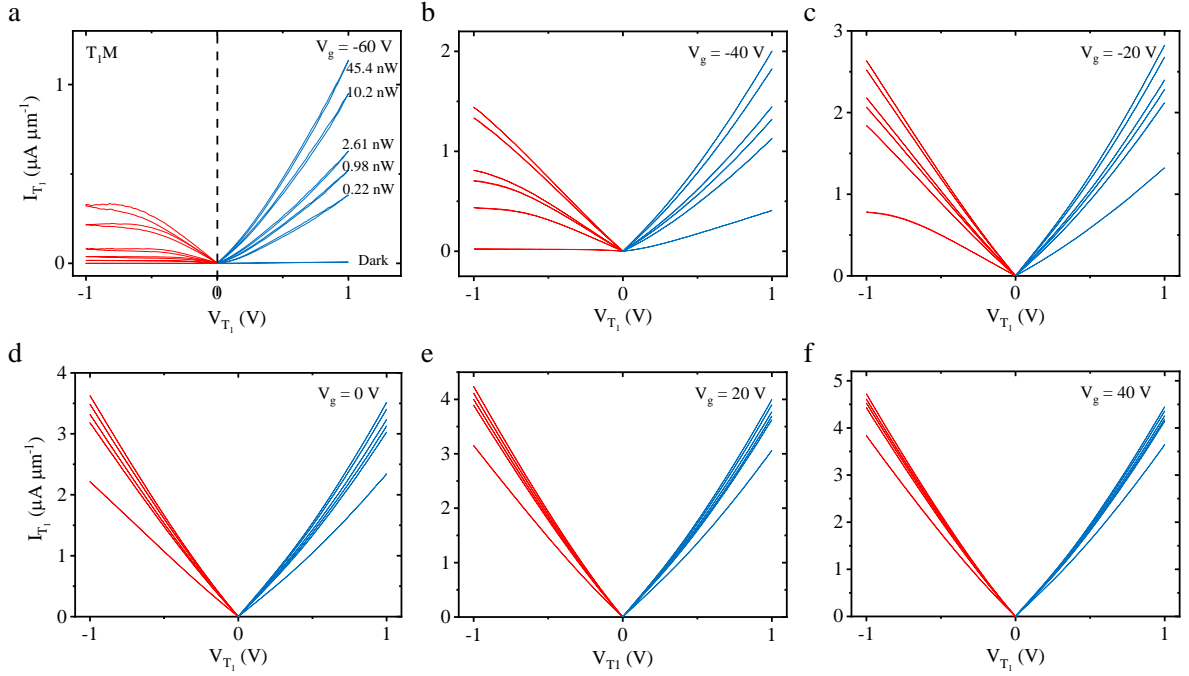

Figure S5: **Light response of heterojunction T$_1$M of device D$_1$.** (a)-(f) Current-voltage characteristics of heterojuntion T$_1$M of device D$_1$ (probed between terminals T$_1$ and M with terminal T$_2$ open) with excitation wavelength of 532 nm at $V_g = -60\ V$ [in (a)], $V_g = -40\ V$ [in (b)], $V_g = -20\ V$ [in (c)], $V_g = 0\ V$ [in (d)], $V_g = 20\ V$ [in (e)] and $V_g = 40\ V$ [in (f)].

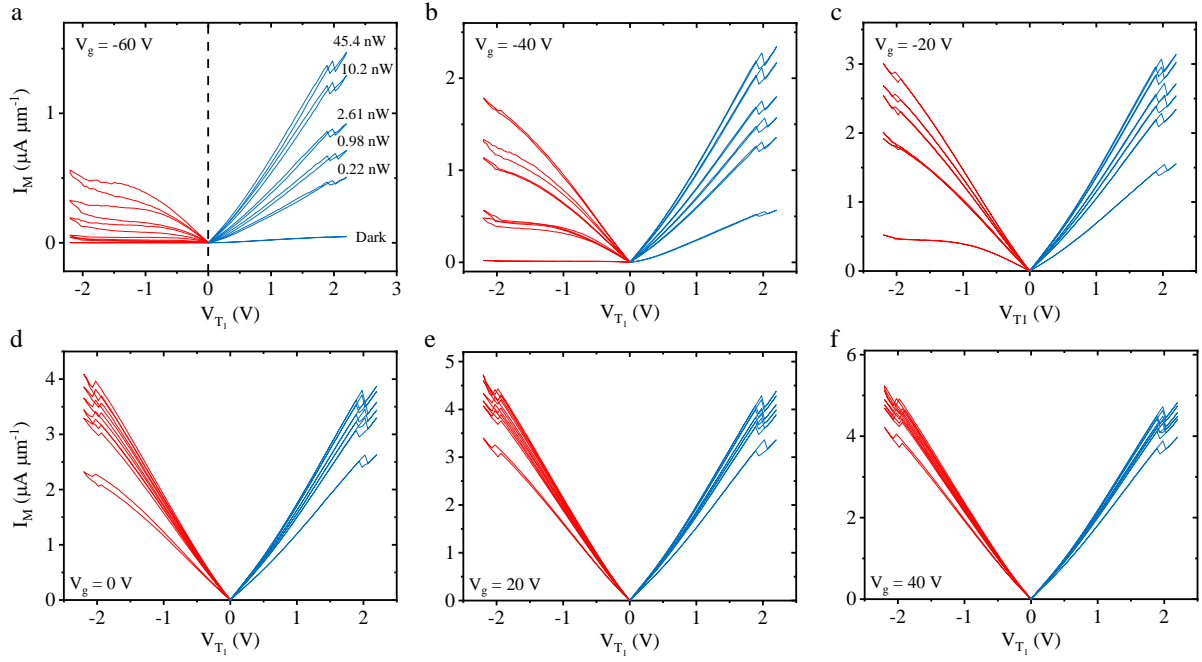

Figure S6: **Light controlled NDR in TaS$_2$/MoS$_2$ asymmetric T-junction (D$_1$).** (a)-(f) $I_M$ *versus* $V_{T_1}$ as the function of 532 nm laser excitation power at $V_g = -60\ V$ [in (a)], $V_g = -40\ V$ [in (b)], $V_g = -20\ V$ [in (c)], $V_g = 0\ V$ [in (d)], $V_g = 20\ V$ [in (e)] and $V_g = 40\ V$ [in (f)].

7

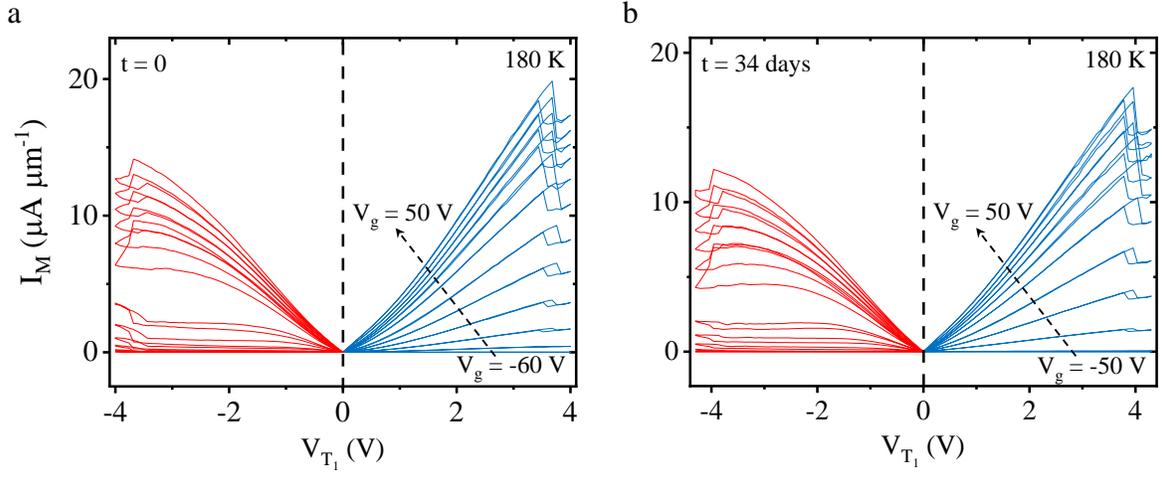

Figure S7: **Stability of TaS$_2$/MoS$_2$ asymmetric T-junction (D$_1$).** (a) $I_M$ *versus* $V_{T_1}$ as the function of back gate voltage ($V_g$) varied from $-60\ V$ to $50\ V$ in steps of $10\ V$ at 180 K. (b) $I_M$ *versus* $V_{T_1}$ depicting NDR after 34 days under similar conditions as (a).

# Graphical TOC Entry

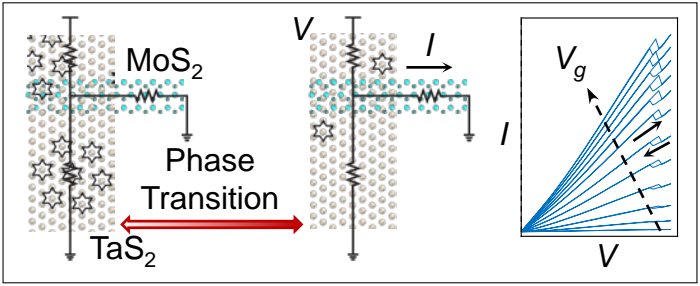


\end{document}